\documentclass[12pt]{article}
 \usepackage[dvips]{graphicx}
\begin{document}
\title {A Note on Transfinite M Theory and the Fine Structure Constant}
\bigskip
\author{~ Carlos Castro \thanks{Center for Theoretical Studies Clark Atlanta
University Atlanta, GA. 30314, USA} }
\date{}
\maketitle
\centerline { Revised , July, 2001}
\begin{abstract}
In this short note , using concepts of $p$-Adic QFT and
$p$-branes, we derive the transfinite $M$ theory  generalization
of the inverse fine structure constant  given by $(\alpha_M)^{-1}
= 100 + 61 \phi $ . The orginal El Naschie and Selvam-Fadnavis
inverse fine structure constant value $(\alpha_{HS})^{-1}= 100 +
60\phi $  was based on a transfinite heterotic string theory and a
quasiperiodic Penrose tiling formalism, respectively. Here $ \phi$
is the Golden Mean $0.6180339...$. Our results are consistent with
the recent astrophysical observations of the Boomerang and Maxima
experiments ,the previous results based on the four dimensional
gravitational conformal anomaly calculations, and with an enhanced
spacetime hierarchy of a suitable number of lines living on Del
Pezzo surfaces.
\end{abstract}
Motivated by the fact that the bosonic membrane is devoid of
anomalies in $d=27 $, and the supermembrane is anomaly free in
$d=11$ [2] , and that the anomaly free ( super) string actions (
$ d =26, 10$ ) are directly obtained by a double-dimensional
reduction process of both the world-volume of the ( super)
membrane and the target spacetime dimension , where the ( super)
membrane is embedded, we shall derive the transfinite $M$
theoretical generalization $(\alpha_M)^{-1} $ to El Naschie's
inverse fine structure constant $(\alpha_{HS})^{-1} $  which was
based on a transfinite Heterotic string theory formalism [1].\\
Selvam and Fadnavis [15]  , independently, obtained this value
using a quasiperiodic Penrose tiling model ( a quasicrystal )
associated with the logarithmic spiral, with a golden-mean winding
number,  which represents a bidirectional vortex ( eddy)
circulation process with a five- fold symmetry. Taking into
account that there are  $5$ steps in this process , both in a
clockwise and counterclockwise motion,  giving effectively a
factor of $ 5 \times 2 = 10$ times the fundamental variance which
in turn is $2 ( 1 +\phi)^4  $ , they obtained the value of the
$total$ variance of the fractal structure associated with the $5$
succesive growth-steps modelled by the Penrose quasi-crystal : $$
Total ~Variance = 5 \times 2 \times 2 ( 1+\phi)^4  = 100 + 60\phi
$$ which agrees exactly with El Naschie's  value based on a
transfinite heterotic string theory formalism. This is no
coincidence. Cantorian fractal spacetime has a $p$-adic topology
and this topology is precisely the one present in those
quasicrystals obtained as limiting  quasiperiodic point sets with
$p$-adic internal symmetries. The Penrose tiling is a special
case of a variety of well known quasiperiodic tilings like the
chair tiling and the Robinson square tiling [17]. Roughly
speaking, the electron is a quasiperiodic process or a
quasicrystal, as we shall see below.\\
We will see as well that the inverse fine structure constant
obtained by the author [ 3] based on a four dimensional
gravitational conformal anomaly calculation [4]
$(\alpha_{conformal})^{-1}$ has for lower/upper bounds the values
$(\alpha_{HS})^{-1}$ and $(\alpha_M)^{-1} $ respectively. All one
needs is to use the following fundamental identities  : $$(
1+\phi)^k = F_{k+1} + F_k \phi.~~~ \phi^k = ( -1)^k F_{k-1} + (
-1)^{ k+1} F_k \phi \eqno ( 1) $$ $$\phi ( 1+\phi ) = 1
\Rightarrow \phi = { \sqrt 5 - 1 \over 2} = 0.6180339....\eqno
(2) $$ where $F_k$ are the Fibonnaci numbers obeying : $$ F_k +
F_{k+1} = F_{ k+2}.~~~F_k =
1,1,2,3,5,8,13,21,34,55,89,144....\eqno (3) $$ $$ \phi = \lim_{k
\rightarrow \infty } ~  { F_k \over F_{k+1}}. \eqno (4) $$ The
identities ( 1 ) can be derived by setting : $$ ( 1 +\phi)^k =
A_k + B_k \phi \Rightarrow (1+\phi)^k + ( 1 +\phi)^{k+1} =
(1+\phi )^{ k+2} = A_{k+2} + B_{k+2} \phi =$$ $$ [ A_k + A_{k+1}
] + [B_k + B_{k+1} ] \phi. \eqno (5) $$ after an induction
process and after using eqs-(3) which define the Fibonnaci
numbers. A $p$-adic QFT argument [5,6,7] allows us to show that
the $M$ theoretical generalization   to the inverse fine structure
constant is  uniquely determined by : $$ (\alpha_M)^{-1} = 1 + (
1+\phi)^2 + ( 1+\phi)^5 + ( 1+\phi)^{10} = $$ $$(F_1 +F_3 + F_6
+F_{11} ) + \phi ( F_2 + F_5 + F_{10} ) = 100 + 61\phi . \eqno
(7a)$$ This expansion of $(\alpha_M)^{-1}$ in $integer$ powers of
$1+\phi $ is the analog of the expansion of $137$ in powers of $
p = 2$ : $$ 137 = 1 + 8 + 128 = 2^0 ( 1 + 2^3 + 2^7 ) \eqno (7b)
$$ and whose $p$-adic norm is $ || 137 ||_2 = (1/2^0 ) = 1 $.
Notice that one could have performed the $lacunar$ series
expansion of the form $$ \sum a_n ( 1+ \phi )^{(1+\phi)^n} \eqno
(7c)$$ a lacunar series is given in powers of the form $ p^{ p^n
} $ but this has $not$ the desired $p$-Adic form because
$(1+\phi)^n $ is not an integer as verified by the fundamental
identities eq-(1). Also we must restrict the coefficients $ a_n $
to be integer valued $ 0,1 < 1+\phi $.\\ El Naschie's value based
on string theory is given by ten copies of the $complexified$
dimensions of the transfinite space ${\cal E}^{ (5) } $ whose
real-valued Hausdorff dimension is $ ( 1+\phi )^4 $ :
$$(\alpha_{HS})^{-1} = 10 \times 2 ( 1+\phi)^4 = 137+ \phi^5 (
1-\phi)^5 = 100 + 60\phi = 137.082.......~~~ \eqno (8 )$$ The
hierarchy of dimensions obtained by El Naschie were given by
suitable powers of $ 10\phi^n $ for $ n = 0, \pm 1, \pm 2, ....$.
The $ n = 0$ power corresponds naturally to the core dimension of
the Heterotic string $ d = 10 $ and the hierarchy of dimensions
for $n = \pm 1,\pm 2$ were $ 4-k ; 6+k; 10; 16+k; 26+k $
respectively, with the value for $ k = \phi^3 ( 1 - \phi^3 )$.\\
However, a close inspection reveals that the quantity $
(\alpha_{HS})^{ -1} = 100 + 60 \phi$ does $not$ admit a $p$-Adic
expansion in the form given by eq-(5a); i.e like the value
$(\alpha_M)^{ -1} = 100+ 61\phi$ does. This is a direct
consequence of the fundamental identities (1). The value of $ 4 -
k $ is not the Hausdorff dimension consistent with the four
dimensional gravitational anomaly calculations [3, 4 ] because
the latter value  is greater than four  . The average dimension
of the world $ d \sim 4 + \phi^3 = 4.236....$ This is another
reason why one should have for core dimension the value of $ d
=11$, the dimension of the anomaly free supermembrane. We will go
back to this crucial point shortly. The value obtained from the
gravitational conformal anomaly calculation was [3] : $$
(\alpha_{conformal})^{-1} = 2 { (\delta^2 + 1)^2 \over \delta (
\delta^2 -1) } = 137.6414382326 . ~~~ D= 4 +\epsilon = 4\delta =
4 + {\phi^3\over 2}.$$ $$\delta = 1 + { \phi^3 \over 8} .~~~\eqno
(9)$$  This value of the inverse fine structure constant is
associated with a fractal spacetime dimension of $ 4 + \epsilon =
4 + ( \phi^3/2) $ . It corresponds to an infrared $fixed$ point
of the renormalization group where scale invariance is restored
[4].\\
 In ordinary ( super ) string theory, a regularization
procedure breaks conformal invariance $except$ in the critical
dimensions $ d = 26; d = 10$ . From the conformal field theory
point of view this implies that the beta function associated with
the worldsheet couplings of the non-linear $\sigma$ model
corresponding to the ( super ) string have a $fixed$ point ; i.e
the beta function vanishes in the critical dimensions and
conformal invariance is restored in those special dimensions.
Hence we have finally : $$ (\alpha_{HS})^{-1} = 100 + 60\phi
=137.082 < ( \alpha_{con})^{-1} =137.6414< (\alpha_M)^{-1} =$$ $$
100 + 61\phi = 137.700\eqno (10) $$ Now we shall go back to the
crucial point of having $d =11$ as the core dimension of the
transfinite nonperturbative $M$ theory. We mentioned earlier that
the bosonic membrane is free of anomalies in $ d = 27 $ and the
supermembrane is anomaly free in $ d = 11$ [2] and that strings
are obtained automatically by dimensional reduction .\\
 The
nonperturbative $M$ theoretic generalization of the inverse fine
structure constant are due to the following $p$-brane
configurations for values of $ p = 0, 2, 5, 10 $ living in $ d
=11$ topological dimensions . The $5$ brane and the membrane , $
p =2$, are Electro-Magnetic dual to each other in $ d =11$ since
EM brane duality requires the numerical relation between the
values of the spatial dimensions to be : $ 11 = 2+5 + 4 $. The $
p = 0$-brane corresponds to the center of mass of the system and
the $ p = 10$ brane is the spacetime filling since $ 10+ 1 = 11$.
Since among our $p$-brane configurations we have the membrane (a
$ p =2$ brane ) living in $ d =11$ dimensions ( and its EM dual
$5$-brane ) this transfinite $M$ theory does naturally $contain$
heterotic strings in $ d =10$ dimensions by a straightforward
dimensional reduction of the membrane.\\ Following Eddington's
view that the inverse fine structure constant is an internal
dimension we can see that the net value $ 100 + 61\phi =
(\alpha_M)^{-1}$ is the sum of the ( fractal ) Hausdorff
dimensions of the four transfinite spaces associated with certain
$p$-branes embedded in $ d =11$ topological dimensions : $$ {\cal
E}^{ (11) } ,~ {\cal E}^{ (6) }, ~{\cal E}^{ (3) }, ~ {\cal E}^{
(1) } \eqno (11)$$ The ( spacetime filling ) $10$-brane spans an
$11$-dimensional worldvolume; the $5$-brane spans a six
dimensional one; its Electromagnetic-Dual Membrane spans a
three-dimensional worldvolume and the center of mass ( $0$-brane)
, spans a one-dimensional line, the normal set $ {\cal E}^{ (1)}
$. These ( topological ) dimensions corresponding to the
worldvolumes, ..... worldlines, spanned by these $p$-branes as
they evolve in time , match precisely the topological dimensions
associated with the four transfinite sets given in eq-(11).\\
 It
is suggestive to think that the four transfinite sets present in
the decomposition (11) could be related to the electromagnetic,
electroweak, strong and quantum gravity phases in Nature. The
number $ 100 + 61\phi $ can be recast explicitly as : $$
(\alpha_M)^{-1} = 137 + \phi^5 ( 1- \phi^5 ) + \phi = 100 +
61\phi . \eqno ( 12)$$ The first term $ 137 $ is the usual $U(1)$
contribution; the second term $ \phi^5 (1- \phi^5 ) = 60\phi - 37
$ is the electroweak and strong interaction contributions found
by El Naschie [ 1] and the $\phi$ additional term is the
Nonperturbative $M$ theoretic corrections; i.e a quantum gravity
effect.\\
 At first hand one may be inclined to $object$ to such an
" erroneous " claim because the contribution of $\phi$ is $not$
small. In fact it is $larger$ than $\phi^5 (1 - \phi^5 ) $.
However, one must realize that in this Nonperturbative
transfinite $M$ theory there is a nontrivial ultraviolet/infrared
entanglement ( mixing ) like that occuring in the construction of
Quantum Field Theories in Noncommutative Geometry . It is
reminiscent of the large-scale/small-scale $T$ duality in string
theory. Therefore, due to this highly nontrivial
ultraviolet/infrared entanglement/mixing the large $\phi$
correction is truly an $infrared$ quantum gravity contribution,
as it should be, since adding this term is precisely what is
needed to $match$ the value obtained from the gravitational
conformal anomaly calculation [ 3,4] of the inverse fine
structure constant. Such contribution was due precisely to an
$infrared$ long distance effect as a result of the quantum
conformal mode fluctuations of the spacetime metric over large
scales whose overall effect is to $screen$ the electric charge at
large distances and cause the $inverse$ value of the fine
structure constant to increase.\\
Within the framework of the the fine structure value  obtained
independently by Selvam and Fadnavis [15]  we can interpret the
additional value of the Golden Mean $\phi$,  to their value of $
100 + 60 \phi$,  due to the intrinsic initial perturbation which
orginated the logarithmic spiraling growth process in succesive
$5$ steps ( clockwise/counterclockwise ).  When one speaks of
growth steps we must specify with respect to what the initial
perturbation of this growth process refers to.\\ Therefore, a
straightforward $p$-Adic QFT argument [5,6,7] yields the
Nonperturbative transfinite $M$ theoretical extensions to El
Naschie-Selvam-Fadnavis inverse fine structure constant. The
value $ 100 + 61\phi$ is consistent with the most recent
Astrophysical data of the spectral density index obtained from the
Boomerang and Maxima experiments measuring the primordial
cosmological background curvature perturbations [8, 9 ] . The
spectral index of the two-point correlation function of the
Cosmic Microwave Background radiation due to the density
perturbations,  induced by the curvature fluctuations, was given
in terms of the quantum trace anomaly [4] and the value is $n
\equiv 2\Delta -3 = 1.0299358 $ where $\Delta = 2.0149679 > 2 $
in full agreement with the numerical bounds of Covi and Lyth
[8].\\
In a forthcoming publication [ 9] we shall present a lengthy and
detailed  discussion of the results discussed in this short note.
Especially , the deep interrelation among $p$-Adic QFT [5,6,7] ;
the Bruhat-Tits tree-like $p$-Adic Topology of Cantorian-Fractal
Spacetime [1] , the Non-Archimedean Geometry associated with the
Algebraic continued-fraction Renormalization group-like procedure
of the physical constants and the $p$-Adic hierarchical
structures of ( symmetry breaking ) phases in
stochastic processes [ 10, 11 ].\\
What remains is to look at the hierarchy of dimensions generated
by this transfinite $M$ theory. The hierarchy generated by the
core dimension of $ d =11 $ is given by $11 \phi^n $ for
different values of $n $ is : $$ n = 0 \Rightarrow d_o= 11 $$ $$
n = - 1 \Rightarrow d_{-1} = 11 ( 1+ \phi ) = 17.7983 $$ $$ n =
-2 \Rightarrow d_{ - 2} = 11 ( 1+\phi)^2 = 11 ( 2+ \phi ) =
28.7983 $$ $$ n = +1 \Rightarrow d_{ + 1 } = 11 (\phi ) = 6.7983
$$ $$ n = + 2 \Rightarrow d_{+2} = 11 (\phi )^2 = 11 - 11\phi
=5-0.7983 = 4.2017 . \eqno ( 13 ) $$ Hence this hierarchy has the
following dimensions : $$ 28.7983, ~17.7983,~ 11, ~ 6.7983, ~ 5-
0.7983= 4.2017  . \eqno ( 14 ) $$ The author has been informed by
Metod Saniga [14] that this hierarchy in its integer valued part
, is almost exactly reproduced by the sequence of number of lines
lying on Del Pezzo surfaces if one adds $one$ to each value : $
28 ; 17; 11; 7; 4 $ that would mean nothing but that Del Pezzo
hierarchy corresponds to spatial dimensions only , while this
enhanced hierarchy grasps both space and time dimensions ! [14].\\
As we said previously, the lower dimension $ 4.2017 $ is
consistent with the four dimensional gravitational anomaly
calculation [3, 4 ] and with the recent Astrophysical data [8, 9].
A  $p$-Adic interpretation to the fundamental scales in Physics
has been given by Pitkanen [7] based on the Mersenne primes : $ M
= 2^m - 1$ for $m$ = prime. Some values of $m$ that yield a
Mersenne prime are : $$ m = 2, 3, 7, 13, 17, 19, 31, 61, 89, 107,
127, 521..... \eqno ( 14 )$$ Pitkanen's  $p$-Adic Length Scale
hypothesis is obtained by selecting the electron Compton
wavelength as a reference scale which allows to fix the size of
an internal $CP^2 $ space whose isometries are linked to the
standard gauge symmetries of the fundamental interactions. The
electron Compton wavelength was chosen to be the one associated
with the Mersenne prime corresponding to $M_{127} = 2^{127} - 1 $
which fixes the size of the internal space to be : $$ l_o \sim
137.6 \times 10^2 ~Planck . \eqno ( 15 ) $$ The hierarchy of
fundamental scales in physics is given by : $$ L = \sqrt {p}  ~
l_o. ~~~ p =M =  2^m - 1 $$ The following values for the Mersenne
primes are :$$1- ~ M_{127}  \Rightarrow L \sim 10^{ - 10} ~cm =
electron $$  $$2- ~ M_{107}  \Rightarrow L \sim 10^{ - 14} ~cm =
proton  $$ $$ 3- ~ M_{89}  \Rightarrow L \sim 10^{ - 16} ~cm = W
~boson $$ $$ 4- ~ M_{61}  \Rightarrow L \sim 10^{ - 20} ~cm =
Michael~Conrad's ~Fluctuon~particle ~[ 20  ]$$  $$5- ~ M_{521}
\Rightarrow L \sim 10^{27 } ~light~years =
Superastronomical~scale $$   $$6- ~ M_{2} = 2^2 -1 = 3 \Rightarrow
L \sim GUT ~scale$$ The reader may ask, do the remaining scales
in between correspond to new particles and forces ? This question
is perfectly valid. For a discussion of $W_\infty $ Geometry and
a master field in $infinite$  dimensions that generates an
infinity of higher spin massless gauge fields ( interactions ) in
lower ( four ) dimensions see [23] .\\
The main conclusion of this work is the following : The value
$(\alpha_{HS} )^{ -1} = 100 + 60 \phi$ does admit the
decomposition $ 10\times 2 ( 1+\phi)^4 $ consistent with the
hierarchy of ( complexified ) dimensions generated by the core $
d = 10 $ dimensions of superstrings but $fails$ to admit the
correct $p$-Adic decomposition exhibited  by $(\alpha_M)^{ -1} =
100 + 61\phi $ shown in eqs-(5).
 Since $61$ is a prime number one can see automatically that
$ 100 + 61\phi $ has for lower and upper bounds the following
 ( complexified ) dimensions belonging to a hierarchy of
dimensions generated by the core $ d = 10, d = 11$ dimensions of
the superstring and the anomaly free supermembrane, respectively :
$$ 10 \times 2 ( 1+\phi)^4  ~ ~ < 100 + 61\phi < ~ 11\times 2 (
1+\phi)^4 . \eqno ( 16 ) $$ Constraining the inverse fine
structure constant to lie in the interval $ 136 < \alpha^{ -1} <
138 $ ; to be  compatible with recent Astrophysical data [8, 9 ]
and with the gravitational conformal anomaly results [3, 4 ] ,
which imposes $ d = 4 + \epsilon $ , we conclude that one cannot
simultaneously satisfy $both$ conditions of $\alpha^{ - 1} $
having the correct $p$-adic expansion like in eq-(5) and in being
a member of the hierarchy of complexified dimensions of the form
$ \alpha^{ -1} = 2d ( 1+\phi)^k$ for $ d= integer$
core-dimensions and $ k =integer $.\\
 One would require to have a relation of the type :
$$ 136 < \sum_n a_n ( 1+ \phi )^n = 2d ( 1+ \phi )^k < 138 .
\eqno (17) $$ it seems that this equation does not have solutions
for $ a_n = 0, 1$ and $d, n, k = integers $ in the range between :
$ 136 < \alpha^{ -1} < 138. $ However, a value of the inverse
fine structure constant like : $ 11 \times 2 (1+\phi)^4 \sim
150.788... $ is $consistent$  with that value obtained by Nottale
using scale relativistic arguments if, and only if,  the number
of Higgs doublets is $7$ [16] . The value obtained by Nottale is :
$$ \alpha^{ -1} = 137.08 + 2.11 ( N_H - 1 ) \pm 0.13 . \eqno (18)
$$ in very good agreement with the experimental value of $137.036$
provided $N_H = 1$ and with the value $11\times 2 ( 1 +\phi )^4
\sim 150.788...$ provided $ N_H = 7 $ .\\

We believe this cannot be just a numerical coincidence but that
an underlying $p$-Adic
internal symmetry operates in the Fractal structures of Nature.\\
T. Smith has used the Hyperdiamond lattice model associated with
the exceptional group $E_8 $,  and Octonions,  to derive the
value of the inverse fine structure constant $ 137.068...$ . In
particular, using the Feynman chessboard construction on
Hyper-Diamond structures gives the physics of the Standard Model
plus Gravitation [18] . Ord has used a similar Feynman chessboard
construction  to formulate a spiral gravity  model [21].  The
$E_8$ lattice can be represented by quaternionic $icosians$ , as
described by  Conway and Sloane [19]. This $E_8$ lattice can be
constructed using  the Golden Mean ratio from the $D_4$ lattice
which has a $24$-cell nearest neighbour polytope. We refer to
[19] for details.\\
Based on the results presented in this work we dare to  say that
the fundamental particles, themselves, like the electron, can be
visualized as a nonlinear dynamical ( multi ) fractal {\bf
process}, represented by a Penrose quasicrystal , for example. It
is this lattice structure that is behind the values of the
fundamental constants in Nature. Multifractality governs also the
prime number distribution [22] .  Not suprisingly, we are going
to see more and more in the near future how all these disciplines
: number theory, algebraic geometry, Cantorian Fractal spacetime,
quasicrystals, $p$-Adic QFT, Noncommutative/Nonassociative
Geometry, fractal strings.... merge together.\\

{\bf{Acknowledgements}}\\
 We are indebted to M. S. El Naschie  and
Metod Saniga for numerous discussions that led to this work.
\bigskip\\
{\bf References }\\
 1- M. S. El Naschie : " A general theory for
the topology of Transfinite Heterotic strings and Quantum Gravity
" Jour. Chaos, Solitons and Fractals {\bf 12 } (5) ( 2001).\\

2- R. Kaiser, U. Marquard, M. Scholl : Phys. Letts {\bf B 227 ]} (
1989 ) 234. U. Marquard, M. Scholl : Phys. Letts {\bf B 227 ]} (
1989 ) 227.\\

3-C. Castro : " On the four-dimensional Conformal Anomaly,
Cantorian-Fractal Spacetime and the Fine Structure Constant " To
appear in Chaos, Solitons and Fractals. Available from
physics/0010072\\

4-I. Antoniadis, P. Mazur, E. Mottola : " Fractal Geometry of
Quantum Spacetime at Large Scales " Available from hep-th/9808070
I. Antoniadis, P. Mazur, E. Mottola " Conformal Invariance and
Cosmic Background Radiation " Available from astro-ph/9611208.\\

5-V. Vladimorov, I. Volovich, E. Zelenov : " p-Adic Analysis in
Mathematical Physics " World Scientific ,
Singapore 1999.\\

6- L. Brekke, P. G. O Freund : " p-Adic Numbers in Physics "
Physics Reports {\bf 233} ( 1) ( 1993)\\

7- M. Pitkanen : " Topological Geometrodynamics : " Book
available on line http:// www.physics.helsinki/$\sim$
matpitka/tgd.htmf\\

8- L. Covi, D. H. Lyth : " Global Fits for the Spectral Index of
the Cosmological Curvature Perturbations "
Available from astro-ph/0008165 v2.\\

9-C. Castro, A. Granik : " In preparation ".\\

10- W. Karkowski , R. Vilela Mendes : "Hierarchical structures
and asymmetric stochastic proceses on $p$-Adics and Adeles "
Jour. Math. Phys. {\bf 35} (9) ( 1994) 4637.\\

11- M. Mezard, G. Parisi, M. Virasoro : " Spin Glass Theory and
Beyond " World Scientifuc, Singapore 1987.\\

12-M. Saniga : " 27 lines on a Cubic Surface and Heterotic String
Spacetimes " Available from physics/0012033. Chaos, Solitons and
Fractals 2001; 12 : 1177-1178\\

13-M. Saniga : ``Lines on Del Pezzo surfaces and transfinite
heterotic string spacetime ``to appear in Chaos, Solitons and
Fractals . Available from physics/0101041.\\

14-M. Saniga : Private Communication . physics/0105049\\

15-A. M. Selvam, S. Fadnavis : ``Superstrings, Cantorian-Fractal
spacetime and quantum like chaos in atmospheric flows ``Chaos,
Solitons and Fractals {\bf 10} (8) ( 1999) 1321-1334.\\

16-L. Nottale : Chaos, Solitons and Fractals 1994, {\bf 4 } ( 3 )
: 361.\\

17 -M. Baake,R. V. Moody, M. Sclottmann : ``Limit quasi periodic
point sets as quasicrystals with $p$-Adic internal spaces
``Available from math-ph/9901008.\\

18-T. Smith : ``From Sets to Quarks ``hep-ph/9708379.\\

 19- J.Conway, N. Sloane : ``Sphere Packings,
Lattices and Groups , 2nd edition, Springer-Verlag ( 1993 ) ``\\

20-M. Conrad  ``Chaos, Solitons and Fractals, 1996  , {\bf 7  } (
5 ) 725.\\

21- G. Ord : The Spiral Gravity Model ``Chaos, Solitons and
Fractals, 1999 , {\bf 10 } (2-3 ) .\\

22- M. Wolf :``Multifractality of Prime Numbers ``Physica {\bf A
160 }( 1989)24.\\

23- C. Castro : ``On the large N limit, $W_\infty$ Strings, Star
products..... ``Available from hep-th/0106260.
\end{document}